\documentclass[epjCONF,columns]{svjour} 
\usepackage{graphics}
\usepackage[varg]{txfonts} 
\usepackage[latin1]{inputenc}
\usepackage{slashed}
\usepackage{enumitem}
\session-title{Hadron Collider Physics Symposium 2011}
\begin{document}
\title{A Novel Technique to Reconstruct the Z mass in WZ/ZZ Events with Lepton(s), Missing Transverse Energy and Three Jets at CDFII.}
\author{Marco Trovato \and Caterina Vernieri\inst{1,2}\fnmsep\thanks{\email{caterina.vernieri@pi.infn.it}}}
\institute{INFN sez. Pisa, Italy \and Scuola Normale Superiore Pisa, Italy }
\abstract{
Observing WZ/ZZ production at the Tevatron in a final state with a lepton, missing transverse energy and jets is extremely difficult because of the low signal rate and the huge background. In an attempt to increase the acceptance we study the sample where three high-energy jets are reconstructed, where about 1/3 of the diboson signal events are expected to end. Rather than choosing the two E$_{T}$-leading jets to detect a Z signal, we make use of the information carried by all jets.\\
To qualify the potential of our method, we estimate the probability of observing an inclusive diboson signal at the three standard deviations level (P$_{3\sigma}$) to be about four times larger than when using the two leading jets only. Aiming at applying the method to the search for the exclusive WZ/ZZ$\rightarrow \ell \nu q\bar{q}$ channel in the three jets sample, we analyzed separately the sample with at least one $b$-tagged jet and the sample with no tags. In WZ/ZZ$\rightarrow \ell \nu b\bar{b}$ search, we observe a modest improvement in sensitivity over the option of building the Z-mass from the two leading jets in E$_T$ . Studies for improving the method further are on-going.
} 
\maketitle
\section{Motivations}
The study of diboson production provides a test of the electroweak sector of the Standard Model (SM). In particular the predicted W$^{\pm}$, Z couplings (Trilinear Gauge Couplings) are sensitive to new physics.\\
The study of associated WZ boson production in the final state $\ell \nu b\bar{b}$ is important since the event topology of this process is the same as expected for WH associated production ($M_{H} \lesssim$ 135 GeV)~\cite{wh}. \\
Observing this process at Tevatron is difficult since the event rate is extremely low. NLO calculations predict WZ production cross section to be $\sim$3.22~pb~\cite{crossWZ}. Thus, one expects a handful of events per fb$^{-1}$ of integrated luminosity in the $\ell \nu q\bar{q}$ final state, after allowing for trigger and kinematical selection efficiency\footnote{
This statement remains valid even if the few accepted ZZ events with leptonic decay of one $Z$, where one lepton is not detected, are included.}. Furthermore, the signal to background ratio is very poor, due primarily to the production of W and associated jets. Since the preferred me- thod used at CDF to disentangle the diboson signal from the backgrounds is a fit to the invariant mass of the two E$_T$-leading jets, an optimal resolution in jet systems mass is of utmost importance.
\section{Three jets}
\begin{figure}\centering

	\resizebox{.95\columnwidth}{!}{\includegraphics{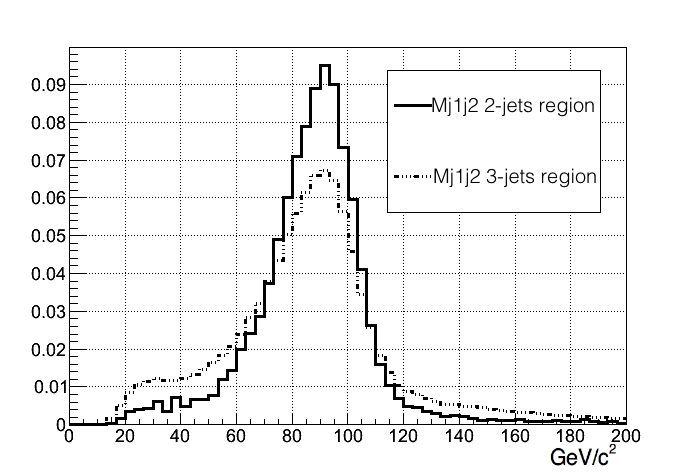} }\\
	\resizebox{.95\columnwidth}{!}{\includegraphics{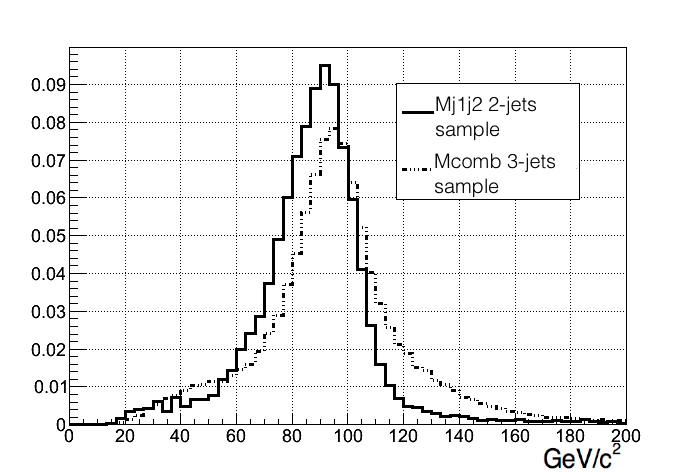} }

	\caption{{Top, MJ1J2 in the \emph{three jets region} (dotted) is compared to MJ1J2 in the \emph{two jets region}. Bottom, MJ1J2 in the \emph{two jets region} is compared to MJJ$_{COMB}$ (dotted) in the \emph{three jets region}. }\label{fig:wrong}
}\end{figure}

In diboson analyses at CDF the standard kinematical cut requires two high energy jets in the candidate sample (\emph{two jets region}). In order to increase signal acceptance, we investigate the sample where three high-E$_T$ jets are found (\emph{three-jets region}), which in simulations is predicted to contain about 33\% of signal events. \\
Additional jets may be initiated by gluons radiated from the interacting partons (Initial State Radiation, ISR) or from the Z-decay quarks (Final State Radiation, FSR)\footnote{
Extra-activity produced by spectator partons or by pile-up of events is negligible in our studies.}. \\
The experimental signature involves the presence of a char-ged lepton (electron or muon), a neutrino (identified through the missing transverse energy, $\slashed{E}_{T}$) and large-$E_T$ jets. \\The sample we investigate is selected by the following cuts:
\begin{itemize}
\item exactly three jets\footnote{Events with a fourth jet with $E_{T} >$~10 GeV are excluded.}~with E$_T({J1,J2,J3}) >$ 25, 15, 15 GeV and $|\eta({J1,J2,J3})| < $2, 2, 3.6\item an isolated triggered electron or muon with $|\eta|<1.1$ and $E_{T}>$ 20 GeV
\item $\slashed{E}_{T}>$ 20 GeV
\item Multi-jet QCD veto:
\begin{itemize}
\item $M^W_T>10\ (30)$ GeV if the triggered lepton is a muon (electron), $M^W_T$ being the $W$-invariant mass in the transverse plane. 
\item  $\slashed{E}_{T}$-significance$\footnote{$\slashed{E}_{T}$-significance~$=(-\log_{10}(\mathrm{P}(\slashed{E}_{T}^{fluct}>\slashed{E_{T}})))$, where P is the probability and $\slashed{E}_{T}^{fluct}$ is the expected missing transverse energy arisen from fluctuations in the energy measurements.}>$~1.8 if the triggered lepton is an electron. 
\end{itemize}
\end{itemize}
In the sample where three jets are found MJ1J2 has a degraded resolution, and high mass and low mass tails due to wrong jet choices are present (see Fig.~\ref{fig:wrong}, top). Our goal is to resolve the combinatorics problem present in this region for building the $Z$ mass and consequently improve the resolution of the invariant mass distribution. This work builds on the efforts of Ref.~\cite{my}.
\subsection{Composition of the selected events}
The following processes would contribute to a data selected sample within our cuts:
\begin{itemize}
\item \textbf{Electroweak and top (EW)}: $WW$, $WZ$, $ZZ$, $Z$+jets, $t\overline{t}$, single-$top$. Each of these processes can mimick the signal signature, with one detected lepton, large $\slashed{E}_{T}$ and jets. The contamination of these processes in the selected data sample is estimated by using their accurately predicted cross sections \cite{crossWZ}. The shapes (templates) of a number of observables are obtained from ALPGEN+Pythia \cite{ALPGEN}, \cite{PYTHIA}, Pythia MC \cite{PYTHIA} after the simulation of the CDF detector.
\item \textbf{W(}$\mathbf{\rightarrow l\nu}$\textbf{)+jets, $l=e,\mu,\tau$}. Due to the presence of real leptons and neutrinos, the $W+jets$ background is the hardest to be reduced. Templates are obtained from ALPGEN+Pythia MC, while the rate normalization is obtained from data.
\item \textbf{QCD}: multi-jet production with a jet faking the lepton and fake $\slashed{E}_{T}$. Since the mechanism for a jet faking a lepton or for fake missing transverse energy is not expected to be well modeled in MC events, both rate normalization and templates are obtained from data. 
\end{itemize}
In Table~\ref{tab:numbers} we show the estimated number of events for each process contributing for the MJ1J2 distribution.

\label{tab:ExpectedPretagEvents}
\begin{table}
\centering                          
\begin{tabular}{lll}
\hline\noalign{\smallskip}
 Process & Rate (Electrons) & Rate (Muons)\\
 \noalign{\smallskip}\hline\noalign{\smallskip}  
Signal & 66.2 $\pm$ 0.9 & 69.5 $\pm$ 0.9\\
$WW$ & 386.2 $\pm$ 3.0 & 311.1 $\pm$ 3.1\\
$t\bar{t}$ & 333.0 $\pm$ 1.4 & 288.5 $\pm$ 1.2\\
single-top & 68.9 $\pm$ 0.4 & 57.8 $\pm$ 0.3\\
$Z$+jets & 350.0 $\pm$ 3.2 & 1167.8 $\pm$ 4.5\\
$W$+jets & 10304.2 $\pm$ 29.6 & 8275 $\pm$ 22.8\\
QCD & 1600.4 $\pm$ 60.0 & 352.3 $\pm$ 5.4\\
\noalign{\smallskip}\hline\noalign{\smallskip}      

Total Observed & 13109.0 $\pm$ 114.5 & 10522.0 $\pm$ 102.6\\
\noalign{\smallskip}\hline
\end{tabular}
\caption[]{Predicted and observed number of events in the notag sample. $W$+jets and QCD rates are estimated by fitting data. The expected rates are separated for different triggered lepton type. By construction the expected numbers are equal to the observed.}
\label{tab:numbers}

\end{table}

\section{Adopted strategy}
\label{StrategyAdopted}
\begin{figure}\centering

	\resizebox{0.465\columnwidth}{!}{\includegraphics{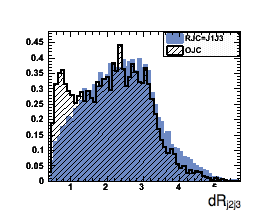} }
		\resizebox{0.45\columnwidth}{!}{\includegraphics{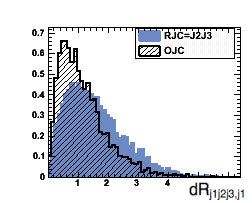} }\\
\resizebox{0.45\columnwidth}{!}{\includegraphics{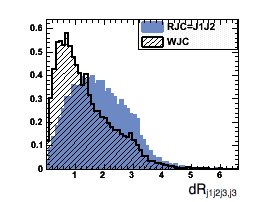} }
	\resizebox{0.45\columnwidth}{!}{\includegraphics{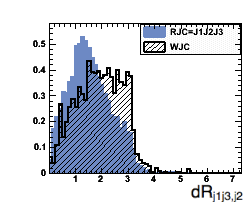} }

	\caption{Some distributions of the variables used as input to NNs,  built for the RJC sample and for the complementary one (shaded).}\label{fig:input}
\end{figure}

We started from studying the three jets sample in WZ MC in which
jets are matched in direction to quarks from Z decay.
Then, we investigate at generator level the origin of the not-matched jet (\textbf{NMJ}) in order to find the Right Jet Combination (\textbf{RJC}) which would give the Z mass. \\
In terms of the RJC frequency the selected sample is composed as follows:
\begin{enumerate}
    \item NMJ = J3 is from ISR $\mapsto$ RJC = J1J2 - 35\% of events
    \item NMJ = J2 is from ISR $\mapsto$ RJC = J1J3 - 21\% of events 
    \item NMJ = J1 is from ISR $\mapsto$ RJC = J2J3 - 10\% of events
    \item NMJ is from FSR $\mapsto$ RJC = J1J2J3 - 19\% of events

  \end{enumerate}
15\% of events cannot be allocated to any of these categories. This problem is a subject of further studies.\\ Four different Neural Networks (NNs) have been trained with the MLP method~\cite{TMVA}, (NN$_{12}$, NN$_{13}$, NN$_{23}$ and NN$_{123}$) in order to decide event by event which RJC should be used. Inputs to NNs are:  
 \begin{enumerate}
 \item Kinematical variables: \\
 d$\eta_{j_{i}j_{k}}$, dR$_{j_{i}j_{k}}$, dR$_{j_{i}\ell}$, dR$_{j_{k}j_{l},j_{p}}$, dR$_{j_{1}j_{2}j_{3},j_{k}}$\footnote{i,~k,~p~$={1;2;3}$. $\ell=$~highest E$_{T}$ lepton}
\item Variables related to the jet systems:\\
-~$m_{j_{i}j_{k}}/m_{j_{1}j_{2}j_{3}}$ \\
-~$\gamma_{j_{i}j_{k}} = (E_{j_{i}}+E_{j_{k}})/m_{j_{i}j_{k}}$ \\
-~$\gamma_{jjj} = (E_{j_{1}}+E_{j_{2}}+E_{j_{3}})/m_{j_{1}j_{2}j_{3}}$ \\
-~`pt-imbalance'~$={P_{T}}_{J1} + {P_{T}}_{J2} - {P_{T}}_{\ell}$- $\slashed{E}_T$\\
-~$\eta(j_{i}+j_{k})/\eta(j_{p})$, $p_{T}(j_{i}+j_{k})/p_{T}(j_{p})$
\item Some tools developed by CDF Collaboration for distinguishing gluon-like and $b$-like jets from light-flavored jets \cite{bness}~\cite{qgluon}.
 \end{enumerate}
 In Fig.~\ref{fig:input} some inputs are shown.\\

\begin{figure}[h!]
	\centering
\resizebox{0.79\columnwidth}{!}{\includegraphics{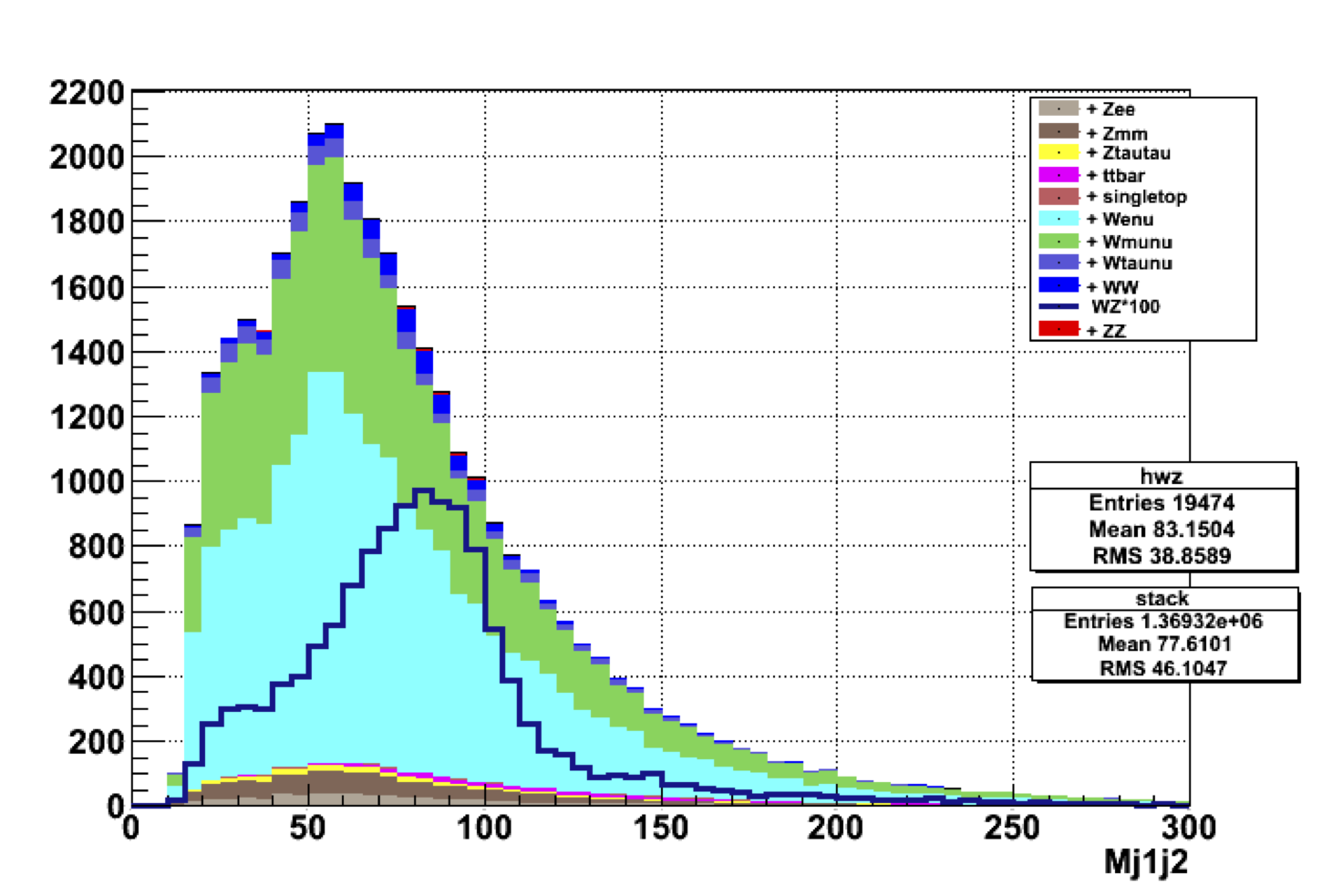}}\\
\resizebox{0.79\columnwidth}{!}{\includegraphics{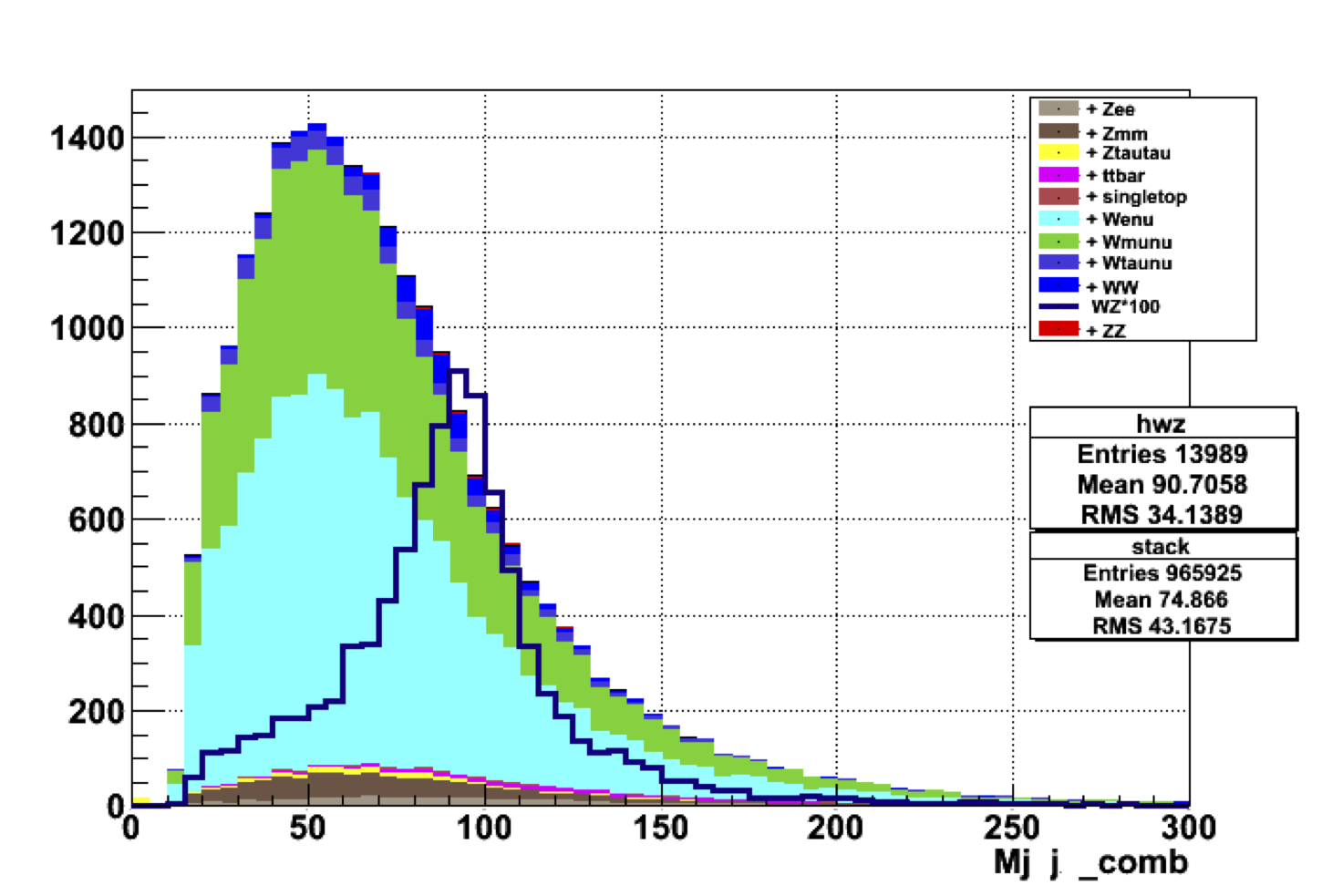}}
  \caption{Simulation of signal+background. Top, MJ1J2. Bottom, MJJ$_{COMB}$. The horizontal scale is in~GeV/c$^{2}$. The signal is multiplied by 100.}
\label{fig:prestack}
    \end{figure}

\noindent 
Combining by a set of subsequent optimal cuts\footnote{Cuts have been optimized against the sensitivity of the measurement.} the information provided by the outputs of the four NNs, we build a ``MJJ$_{COMB}$" Z-mass. Using MJJ$_{COMB}$ rather than MJ1J2, resolution improves by a factor $\sim$2, see Fig.~\ref{fig:wrong} and Table~\ref{preresul}. \\
We apply the method also to the major sources of background of a typical diboson analysis at CDF~(W+jets,  Z+ jets, $t\bar{t}$ and single top) and compare the result to WZ events. In Fig.~\ref{fig:prestack} and in Table~\ref{preresul} can be noticed that MJJ$_{COMB}$ allows a better separation of the WZ/ZZ signal from background.
\begin{table}
\centering                          
\begin{tabular}{lll}
\hline\noalign{\smallskip}
   & MJ1J2& MJJ$_{COMB}$\\
\noalign{\smallskip}\hline\noalign{\smallskip}                        
$Acc$ &{100\%} &{72\%}\\
$p$ &{35\% }&{ 64\%}\\
$\sigma/\mu$ & 0.27 & 0.13\\
\noalign{\smallskip}\hline
\end{tabular}
\caption{\emph{Acc} is the acceptance; $p$ is the purity and it is defined as the fraction of events where the corrected jets are selected; $\sigma$ and $\mu$ are estimated by a Gaussian fit in the mass window $[$70,110$]$ GeV/c$^{2}$}.
\label{preresul}
\end{table}
\section{Tests of the method}
To qualify the potential of the method we have studied an experimental data sample accepting events with an isolated large E$_T$~(p$_T$) lepton, large missing E$_T$ and three large transverse momentum jets. The selection cuts accept jets of all flavors (\emph{pretag} sample), and all diboson events including WW besides WZ, ZZ may pass the cuts. We estimate the probability at three standard deviations level to extract an inclusive diboson signal. After our procedure for building the Z mass is applied, P$_{3\sigma}$ is about 4 times greater than when building the Z mass ``by default'' with the two E$_{T}$ leading jets.\\
This attempt represents just a check of our technique. Since diboson signal has been observed in CDF~\cite{vivi}, it would be an useful test to understand if using only the three-jets sample a diboson\footnote{we expect the $ZZ$ contribution to be negligible due to the requirement on $\slashed{E_{T}}$} signal could be extracted.\\ 
In order to discriminate WZ against the WW contribution we apply our technique considering only WZ/ZZ as the signal. We decide to treat separately the \emph{notag} and \emph{tag} three jets regions and then combine the results in order to reach a greater sensitivity. The sensitivity increases when MJJ$_{COMB}$ rather than the standard MJ1J2 is used: the expected $p$-value is about 30\% greater in the former case. \\
In conclusion, our technique allows including the three jets sample in the WZ/ZZ search in order to increase acceptance and sensitivity in the search for the hadronically decaying Z-boson. \\
Improvements to this technique and other possible applications are being investigated.
\begin{table}
\begin{center}\begin{tabular}{lll}
\hline\noalign{\smallskip}
{Fit Method}&${P_{2\sigma}}$&${P_{3\sigma}}$\\
\noalign{\smallskip}\hline\noalign{\smallskip}
Fit signal~$WZ/ZZ/WW$~(pretag) &&\\
- MJ1J2 &51.2\%&6.4\%\\
- MJJ$_{COMB}$  &66.7\%&25.9\%\\

\noalign{\smallskip}\hline\noalign{\smallskip}
&$p$-value&\\
\noalign{\smallskip}\hline\noalign{\smallskip}

Fit signal~$WZ/ZZ$~(notag+tag)&&\\
- MJ1J2 &0.35 $\sigma$&\\
- MJJ$_{COMB}$  &0.45 $\sigma$&\\
\noalign{\smallskip}\hline

\end{tabular}
\end{center}\caption{Sensitivity of the fits considering only the three jets region.}
\end{table}

\begin{acknowledgement}
We are grateful to Prof. Giorgio Bellettini, Dr. Giuseppe Latino and Dr. Vadim Rusu for many fruitful discussions and suggestions.
\end{acknowledgement}

%


\begin{thebibliography}{}
\bibitem{wh} CDF Collaboration [arXiv:1112.4358v1] (2011)
\bibitem{crossWZ} J. M. Campbell and R. K. Ellis, Phys. Rev. D \textbf{65}, 113007 (2002).

\bibitem{my} M. Trovato, C. Vernieri, J Phys. Conf. Series \textbf{323}, 012014 (2011)

\bibitem{ALPGEN} M. Mangano, M. Moretti M, F. Piccinini, R. Pittau and A. Polosa, {J. High Energy Phys.} \textbf{07}, 001 (2001)
\bibitem{PYTHIA} T. Sj\"ostrand, P. Ed\'en, C. Friberg, L. L\"onnblad, G. Miu, S. Mrenna and E. Norrbin  {Computer Phys. Commun.} \textbf{135}, 238 (2001).

\bibitem{bness} J. Freeman, W. Ketchum, J.D. Lewis, S. Poprocki, A. Pronko, V. Rusu, P. Wittich, {Nuclear Instruments and Methods in Physic Research~Sec.~A}~\textbf{663}, 37 (2012)
\bibitem{TMVA} A. Hoecker et al. \emph{TMVAUsersGuide} ({http: tmva. sourceforge.net}, 2009)

\bibitem{qgluon} W. Ketchum, CDF Public Note \textbf{10643} (2011)
\bibitem{vivi}T. Aaltonen et al. , Phys. Rev. Lett. \textbf{104}, 101801 (2010)
\end{thebibliography}
\end{document}